\documentclass{PoS}
\usepackage{epsfig}

\PoS{PoS(LAT2005)134}

\title{Low Energy Constants from the zero mode contribution to the pseudo-scalar correlator}

\ShortTitle{LEC from the zero mode contribution to the
pseudo-scalar correlator }

\author{\speaker{Stanislav Shcheredin}\thanks{HU-EP-05/43, SFB/CPP-05-45, BI-TP 2005/32}\\

Fakult\"at f\"ur Physik\\
Universit\"at Bielefeld\\ Universit\"atsstra\ss e\\ 
D-33615 Bielefeld, Germany\\

E-mail: \email{shchered@physik.hu-berlin.de}}

\author{Wolfgang Bietenholz\\

Institut f\"ur Physik\\ 
Humboldt Universit\"at zu Berlin,\\  
Newtonstr.\ 15\\
D-12489 Berlin, Germany\\

E-mail: \email{bietenho@physik.hu-berlin.de}}

\abstract{We apply different types of overlap operators
in quenched QCD simulations to compute the zero mode contribution 
to the pseudo-scalar correlator. In
particular we use the conventional Neuberger Dirac operator and
the overlap hypercube Dirac operator. Confronting our data with
the analytical predictions by Chiral Perturbation Theory we
evaluate the pion decay constant and the parameter $\alpha$ of the
quenched chiral Lagrangian.}

\FullConference{XXIIIrd International Symposium on Lattice Field Theory\\

         25-30 July 2005\\

         Trinity College, Dublin, Ireland}

\begin{document}

\section{Introduction}
Chiral Perturbation Theory ($\chi$PT) has proven to be a
valid theoretical framework for the light mesons at
small energies, less than the chiral symmetry breaking scale $4\pi
F_{\pi}$, where $F_{\pi}$ is the pion decay constant. For 
$3$-flavour QCD the mesons belong to the coset space of $U\in
SU(3)\otimes SU(3)/SU(3)$. The chiral Lagrangian is constructed as
a hierarchical expansion in two sets of parameters, the meson
momenta and the quark masses. This has to be compatible with the
underlying global chiral symmetry of QCD. To render the theory a
one parameter expansion one adopts a counting scheme. In the
leading order, the chiral Lagrangian is parameterised by the two
low energy constants (LEC) $F_\pi$ and $\Sigma$,
\begin{equation}
{\cal L}^{(2)}[U] = \frac{F^2_{\pi}}{4}{\rm Tr}(\partial_\mu U
\partial_\mu U^\dagger ) -\frac{\Sigma}{2}{\rm
Tr}(MU^\dagger + {\rm h.c.}) \ , \quad M={\rm diag}(m_u,m_d,m_s) \ ,
\end{equation}
where $\Sigma$ is the chiral condensate in the
chiral limit. At zero quark mass the pion decay
constant is expected to be $86$~MeV~\cite{CD} whereas its physical
value is $93$ MeV. A conventional counting scheme, where one takes
the pion mass ($m_\pi \simeq \sqrt{2m_q \Sigma/F^2_\pi}$, with 
$m_q = m_u = m_d$) to be of the
same order as the pion momentum, fully determines the hierarchy of
the chiral expansion.

A theoretical LEC determination is only possible
with the aid of ab initio QCD calculations. 
The aim of our ongoing project is to explore the
feasibility of a LEC determination by examining different
observables and lattice fermions, see also
Ref.~\cite{Bie0} for another contribution to these proceedings.

In a finite volume $a^4L^3\times T$ one distinguishes between two
expansions of the $\chi$PT: the so-called
$p$-regime and the $\epsilon$-regime. The $p$-regime \cite{preg}
is defined in a way that the pion fits well in this box, i.e.
\begin{equation}
L\gg \frac{1}{m_\pi} \ .
\end{equation}
Here one applies the conventional counting scheme to define the
chiral expansion. In the $p$-regime the quantities
receive finite size effects which fall off as \ ${\rm
exp}{(-m_\pi L)}$.

In the $\epsilon$-regime \cite{eps1} one squeezes the pion into a small 
box and tries to calculate the resulting finite size effects. Here the
following inequalities have to hold
\begin{equation}
 \frac{1}{4\pi F_\pi} \ll L< \frac{1}{m_\pi}\, .
\end{equation}
The first condition ensures that the typical momentum in a box is
much smaller than the chiral symmetry breaking scale, i.e. the
expansion parameter is small. 
In this setting one has to change the counting scheme as
\begin{equation}
 \left(\frac{p}{4\pi F_\pi}\right)^2 \sim \frac{m_\pi}{4\pi
 F_\pi} \ .
\end{equation}
Since the LEC entering the expressions for the
observables in the $\epsilon$-regime are those of the infinite
volume, one can obtain them by controlling the finite size effects
in the $\epsilon$-regime. In this setting one can perform
simulations with extremely light quarks without the requirement of
a huge volume, hence the $\epsilon$-regime appears very appealing
for lattice simulations. Unlike the $p$-regime, the observables
in the $\epsilon$-regime strongly depend on the topological
sector~\cite{eps2}.
Therefore one of the important properties of the lattice fermion
formulation is that it must give a sound definition for the
topological charge. This is fulfilled if the lattice Dirac
operator $D$ obeys the Ginsparg-Wilson relation~\cite{GW}
\begin{equation}
\{D,\gamma_5\}=\frac{a}{\rho} D\gamma_5D \ .
\end{equation}
It provides us with a formulation of lattice
fermions with exact lattice chiral symmetry~\cite{Luscher}. The
topological charge is identified with the index $\nu$ of this operator,
which has exact left-handed and right-handed zero
modes~\cite{Has}. A particular solution to the Ginsparg-Wilson
relation is given by the massless Neuberger operator~\cite{NEU}
\begin{equation}
D_{\rm ov}=\frac{\rho}{a}\left[ 1+A/\sqrt{A^\dagger A} \ \right] \
, \quad A=a\hat{D}-\rho \, , \quad \hat{D}=D_{\rm W} \ ,
\end{equation}
where the Wilson Dirac operator $D_{\rm W}$ is inserted into the overlap
formula with a negative mass term $\rho/a$. Pursuing better
locality, rotational symmetry and scaling properties, one is
motivated to use a different kernel in the overlap
formula~\cite{Bie1}. As an alternative to the Wilson kernel
we employed the hypercube operator~\cite{Bie2}. Its
construction is based on a truncated free perfect fermion
gauged with hyperlinks. This operator has the structure
\begin{eqnarray}
D_{\rm HF}(x,y) = \sum_{\mu=1}^4\rho_\mu(x-y;U)\gamma_\mu  
+ \lambda(x-y;U) \ ,
\end{eqnarray}
where $x-y$ is restricted to a unit hypercube. We improved the
chiral properties of this kernel by amplification factors for the
couplings of the free perfect fermion and by fattening the links
before constructing the hyperlinks~\cite{Bie3}. The resulting
operator is called the hypercube operator and when inserted in the
overlap formula we obtain the overlap hypercube operator. Since
the kernel has already good chiral properties it is expected that
the overlap formula corrects it only mildly. Thus the main virtues
of the truncated perfect action are preserved to a large extent.

Due to the substantial computational overhead of the overlap operator over 
the non-chiral operators, its applications are still mostly limited to
the quenched approximation. To interpret the results of quenched
QCD simulations, quenched $\chi$PT has been
constructed~\cite{Quen}. A down side of the quenched computations
is that they suffer from logarithmic volume corrections;
for instance $\Sigma$ is not well defined in the infinite volume
limit~\cite{logq}.

In this study we concentrate on a quenched lattice QCD computation
of the time derivative of the zero mode contribution to the
correlator of the pseudo-scalar density $P = \bar \psi \gamma_5 \psi$,
\begin{equation}
C'_{|\nu|}(t)=\frac{d}{dt}\sum_{\vec{x},i,j}\Big\langle \langle
j,\vec{x}, t|i,\vec{x},t \rangle \langle i,0|j,0\rangle
\Big\rangle_{\nu} \quad , 
\qquad D_{\rm ov} \, |i,\vec{x},t\rangle=0 \ .
\end{equation}
This quantity has been computed analytically to NNLO in $\chi$PT
as well as quenched $\chi$PT in Ref.~\cite{zero}, which also
presented a numerical study with the Neuberger operator. We extend
this study to the overlap hypercube operator.

The quenched chiral Lagrangian to NNLO reads
\begin{equation}
{\cal L^{\rm NNLO}}=\frac{F^2_{\pi}}{4}{\rm Str}(\partial_\mu U
\partial_\mu U^\dagger ) -\frac{\Sigma}{2}{\rm
Str}(MU + {\rm h.c.}) -iK\Phi_0{\rm Str}(MU -{\rm h.c.})
+\frac{m^2_0}{2N_c}\Phi^2_0+ \frac{\alpha_0}{2N_c}(\partial_\mu
\Phi_0)^2\, ,
\end{equation}
where $U \in SU(N_f|N_f)\otimes SU(N_f|N_f)/SU(N_f|N_f)$ and 
``Str'' is the corresponding ``supertrace''.\\
${\cal L^{\rm NNLO}}$ contains now five quenched LEC: 
$F_\pi$, $\Sigma$, $K$, $m_0$, $\alpha_0$. 
The peculiarity of the quenched chiral
Lagrangian is the presence of a scalar field $\Phi_0$, which is
responsible for the quenching effects.

We consider only the first term in the Taylor expansion of
$C'_{|\nu|}(t)$ about $T/2$. In a box $L^3\times T$ it takes
the form~\cite{zero}
\begin{equation}
C'_{|\nu|}(t)/L^2=D^{\rm NNLO}_{|\nu|}(F_\pi, \langle
\nu^2\rangle,\alpha)\frac{s}{T} +{\cal O}\left ( \left
(\frac{s}{T}\right )^3\right )\, , \quad s=t-T/2 \ ,
\label{Taylor}
\end{equation}
where $\alpha$ is defined by \ $\alpha/2N_c := \alpha_0/2N_c
-2KF_\pi/\Sigma \,$, and the dependence on $\langle \nu^2\rangle$
originates from the Witten-Veneziano formula~\cite{WV}.

\section{Numerical setup and results}

\begin{figure}
\hspace{-0.7cm}
\includegraphics[width=0.44\textwidth,angle=-90]{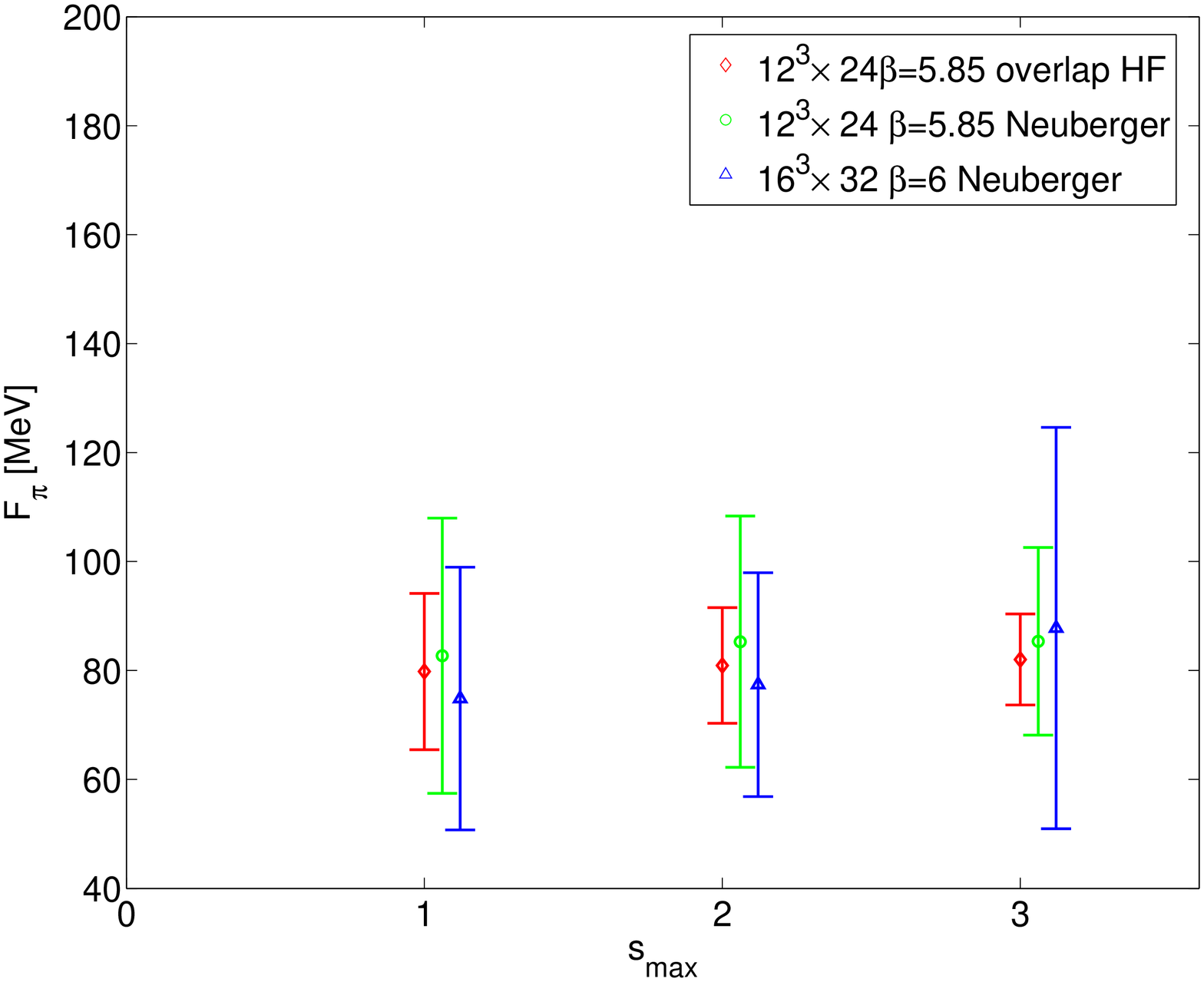}%
\hspace*{1mm} 
\includegraphics[width=0.44\textwidth,angle=-90]{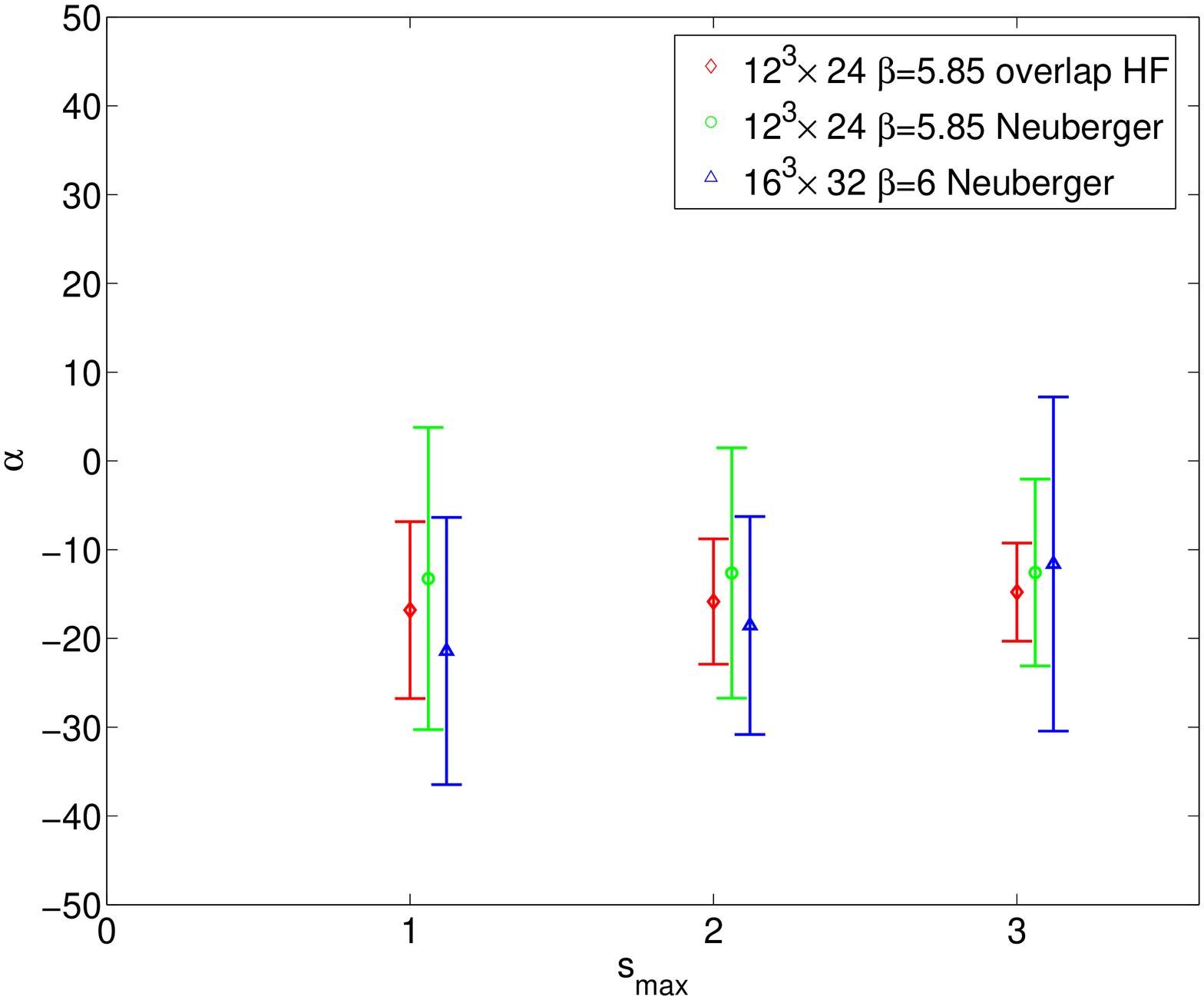}
\caption{$F_\pi$ and $\alpha$ according to 
Eq.~(1.10) vs.\ the width of the
fitting range $s_{\rm max}$.} \label{fig1}
\end{figure}

We performed quenched QCD simulations with the Wilson gauge action
at $\beta=5.85$ on a lattice of size $12^3\times 24$, and at
$\beta=6$ on $16^3\times 32$. The parameters for the
overlap operators are described in Ref.~\cite{preliminary}. In
both cases, the physical volume amounts to
$(1.48\,{\rm fm})^3\times 2.96 \,{\rm fm}$.  
We chose it large enough, referring to the
experience with the chiral Random Matrix Theory~\cite{RMT} and to
the behaviour of the axial correlators in the
$\epsilon$-regime~\cite{axial}. We computed zero eigenmodes for
the overlap hypercube operator at $\beta=5.85$ and for the Neuberger 
operator on both lattices. Our statistics is collected in Table~\ref{tab1}.
\begin{table}
\begin{tabular}{|l|c|c|c|c|c|c|c|c|}
\hline
 & $\rho$ & $\langle \nu^2 \rangle$ &&
 $|\nu|=1$ & $|\nu|=2$ && $F_\pi$ [MeV] & $\alpha$ \\
 \hline
overlap HF $12^3\times 24$, $\beta=5.85$ & 1 & $10.8$ &&
$221$ & $192$ && $80 \pm 14$ & $-17 \pm 10 $ \\
 \hline
Neuberger $12^3\times 24$, $\beta=5.85$ & 1.6& $10.7$ &&
$132$ & $115$ && $83 \pm 25$ & $-13 \pm 17$ \\
 \hline
Neuberger $16^3\times 32$, $\beta=6$ & 1.6& $10.4$ &&
$115$ & $94$ && $78 \pm 30$ & $-19 \pm 18$ \\
 \hline
\end{tabular}
\caption{Statistics and results for $F_\pi$ and $\alpha$ at
fitting range $s_{\rm max}=1$.} \label{tab1}
\end{table}
We fixed the value of $\langle \nu^2\rangle$ from our simulations
and performed a combined two parameter fit of the leading term in
Eq.~(\ref{Taylor}) to our data in the
topological sectors $|\nu|=1,2$ in intervals $s \in
[-s_{\rm max} ,s_{\rm max}]$, with
$s_{\rm max}=1 \dots 3$. Preliminary results were
presented in Ref.~\cite{preliminary}, and our new results are
given in Figure~\ref{fig1} and in Table~\ref{tab1}. 
Our values for $F_{\pi}$ and $\alpha$ are a little lower than 
the results reported in Ref.\ \cite{zero}
(for the Neuberger operator in isotropic boxes).\footnote{A 
different way to compute $F_\pi$ in the
$\epsilon$-regime is to use the axial-vector
correlator~\cite{Bie0,axial,Fukaya:2005yg}.
That method suggests a somewhat larger value of $F_{\pi}$
(at finite $m_{q}$).}

We still have sizable (jackknife) error bars.
We see, however, that our values of
$F_\pi$ and $\alpha$ are compatible within the errors for
different overlap fermions and for two lattice spacings.
In particular the result with the overlap hypercube operator 
favours a negative value for the parameter $\alpha$. 
A negative value has also been obtained
for the ``original'' parameter $\alpha_0$ 
from the full pseudo-scalar correlator \cite{Fukaya:2005yg}.

\section{Conclusions and outlook}
We computed quenched values of $F_\pi$ and the parameter $\alpha$
from the zero mode contribution to the pseudo-scalar correlator
using the Neuberger and the overlap hypercube operator 
in the chiral limit, in a volume 
$(1.48\,{\rm fm})^3\times 2.96 \,{\rm fm}$.
The results are compatible for two lattice spacings and
different overlap operators within the errors. 
Our result for $F_{\pi}$ obtained with this method 
(in the given volume) is close to the
phenomenological value (in the chiral limit).
It would be interesting to extend this
study to a finer lattice spacing for the overlap HF and to
increase our statistics. 

\acknowledgments We would like to thank
M.\ Papinutto and C.\ Urbach for providing us with powerful
eigenvalue routines. We acknowledge discussions with S.\ D\"urr,
K.-I.\ Nagai, H.\ St\"uben and P.\ Weisz. This work is supported by
the Deutsche Forschungsgemeinschaft through SFB/TR9-03. The
computations have been performed on the IBM p690 clusters of the
HLRN ``Norddeutscher Verbund f\"ur Hoch- und
H\"ochstleistungsrechnen'' (HLRN) and at NIC, Forschungszentrum
J\"ulich.

\end{document}